\documentclass[%
jmp,%
rsi,%
 amsmath,amssymb,
preprint,%
]{revtex4-1}

\usepackage{graphicx}
\usepackage{dcolumn}
\usepackage{bm}
\usepackage{float}
\usepackage{color}
\def\FH{\textcolor{black}}

\begin{document}


\title[KEDF Expansion]{Kinetic Energy Density Functionals by Axiomatic Approach}

\author{Fahhad H. Alharbi}
 \email{falharbi@hbku.edu.qa}
 \affiliation{Qatar Environment and Energy Research Institute (QEERI), Hamad Bin Khalifa University, Doha, Qatar.}
 \affiliation{College of Science \& Engineering (QEERI), Hamad Bin Khalifa University, Doha, Qatar.}
 
\author{Sabre Kais}
 \email{kais@purdue.edu}
 \affiliation{Qatar Environment and Energy Research Institute (QEERI), Hamad Bin Khalifa University, Doha, Qatar.}
 \affiliation{College of Science \& Engineering (QEERI), Hamad Bin Khalifa University, Doha, Qatar.}
 \affiliation{Department of Chemistry, Physics, and Birck Nanotechnology Center, Purdue University, West Lafayette, Indiana 47907, USA.}

\date{\today}

\begin{abstract}
An axiomatic approach is herein used to determine the physically acceptable forms for general $D$-dimensional kinetic energy density functionals (KEDF). The resulted expansion captures most of the known forms of one-point KEDFs. By statistically training the KEDF forms on a model problem of non-interacting kinetic energy in 1D (6 terms only), the mean relative accuracy for 1000 randomly generated potentials is found to be better than the standard KEDF by several orders of magnitudes. The accuracy improves with the number of occupied states and was found to be better than $10^{-4}$ for a system with four occupied states. Furthermore, we show that free fitting of the coefficients associated with known KEDFs approaches the exactly analytic values. \FH{The presented approach can open a new route to search for physically acceptable kinetic energy density functionals and provide an essential step towards more accurate large-scale orbital free density functional theory calculations.}
\end{abstract}

\maketitle

\section*{Introduction}

Presently, density functional theory (DFT) dominates the field of atomistic and molecular quantum chemistry calculations \cite{J01,B01,K01}. This is mainly due to its relatively low computational cost compared to many other atomistic approaches. The mainstream DFT (i.e. Kohn-Sham DFT (KS-DFT) \cite{K02}) is a slight alteration of the original work of Hohenberg and Kohn (HK-DFT) \cite{H01} where it was proved that the ground state of any many-electron system is completely characterized by its density and the energy functional which permits the system to attain its minimum at the density corresponding to the ground state. However, representing the contribution of the kinetic energy as a density functional (KEDF) \FH{($T\left[ \rho\left( \textbf{x} \right) \right] = \int t \left( \textbf{x} \right) d\textbf{x}$) where $t \left( \textbf{x} \right)$ is the kinetic energy density (KED)} has proven to be challenging as the accuracy and applicability of the presently proposed KEDF are generally not sufficient for reliable calculations \cite{K01,P01,C01,A01,FH3,FH4}. So, Kohn and Sham (KS-DFT) suggested an approximate approach where the ``orbitals'' are reintroduced such that the sum of the orbitals densities equals to the exact density of the real system and the kinetic energy is defined as the kinetic energy of the introduced ``fictitious'' system. Computation-wise, this results in a conversion of the problem from 3-dimensional (3D) to $3N$-dimensional where $N$ orbitals are determined by solving the governing $N$-3D equations self-consistently and $N$ is the number of electrons in the calculations.

Recently, the search for an \textit{orbital-free} version of DFT (OF-DFT) has rapidly gained attention \cite{C02,E01,K03,G02,C04}. To apply ``orbital free'' approaches, it is essential to find highly accurate kinetic energy density functionals (KEDF). The subject is not new and its origins date back to the early years of quantum mechanics. Most of the early proposed KEDF are based on particular exactly solvable models, e.g. constant potential with plainwaves solution for Thomas-Fermi model (TF) \cite{T01,F01} or modified planewaves for von Weizsacker KEDF (vW) \cite{V01}. This period was followed by many extensions and  further developments on ``specific physical models''. For a comprehensive collection of suggested functionals, we refer the reader to a review by Tran and Wesolowski \cite{T02}. Alternatively, some of the recent developments adopted statistical techniques. For example, Burke and coworkers recently used machine learning to approximate density functional \cite{S01,S02} based on statistical expressions.

In this paper, we revisit this from a different perspective. Instead of starting from a specific physical model or using mathematical tools to improve a KEDF, we begin by asking which forms of KEDF are physically acceptable. To answer this query, an axiomatic approach is used to determine the physically acceptable forms in a general $D$-dimension space. In axiomatic approaches, a set of axioms are used in conjunction to derive equations, expressions, or theorems.

As expected, the resulted expansion captures most of the known forms of one-point KEDFs. By statistically training the forms for a model problem of the non-interacting kinetic energy in 1D (6 terms only), we find that the mean relative accuracy for 1000 randomly generated potentials is orders of magnitudes better than that delivered by standard KEDFs. The accuracy improves with the number of occupied states, and it is better than $10^{-4}$ at four occupied states. Furthermore, it is shown that the free fitting of the coefficients associated with known KEDFs approaches the exactly known values.

\FH{
\section*{Kinetic Energy Density Expansion for $D$-Dimensional Space}
}

Herein, an axiomatic approach is used to derive an expansion of physically acceptable forms of KEDF that satisfy the following essential physical requirements; dimensionality, finiteness, compatibility with the virial theorem, and non-negativity of $t \left( \textbf{x} \right)$ . Also, the functional derivative of the final expansion will be derived as it is needed for efficient energy minimization.

\subsection*{The expression of $t \left( \textbf{x} \right)$ as a function of density derivatives}

Generally, $t \left( \textbf{x} \right)$ can be defined as a scalar function of the density and its derivative, i.e.
\begin{equation}
\label{GEF}
	t\left( \textbf{x} \right) \equiv f \left( \rho , \nabla \rho , \nabla^2 \rho , \dotsm , \nabla^{(n)} \rho \right) \, .
\end{equation}
For the 1-dimensional case (1D), Shao and Baltin \cite{S03} proved that the highest derivative order is 1 and all the derivatives for $n>1$ must be ruled out based on the compatibility with the differential virial theorem \cite{R01,H02,H03} and $t \left( \textbf{x} \right)$ non-negativity . The same concept can be further extended to $D$-dimensional cases.

\FH{Here, we outline Shao and Baltin \cite{S03} proof for 1D case and how it is extended to $D$-dimension. In 1D, the differential virial theorem of fermionic system with Coulombic interaction is
\begin{equation}
\label{DVT1D}
	\rho(x) v'(x) = \dfrac{1}{4} \rho''' - 2 \, t'(x) \, .
\end{equation}
At the beginning, we need to represent the above equation in terms of the density and its derivatives only. $t'(x)$, which is directly obtained from Eq. \ref{GEF} (for 1D), is
\begin{equation}
\label{R2}
	t'(x) = \sum_{\nu=0}^n \rho^{(\nu+1)} \frac{\partial f}{\partial \rho^{(\nu)}}
\end{equation}
As for $v'(x)$, it is obtained by taking the derivative of Euler equation when applying Eq. \ref{GEF} within DFT framework. So,
\begin{equation}
	\frac{d}{dx}\left( \frac{\delta E[\rho]}{\delta \rho} \right)=0
\end{equation}
and hence
\begin{equation}
\label{R3}
	v'(x) = - \frac{d}{dx} \left( \frac{\delta E[\rho]}{\delta \rho} \right) = \sum_{\nu=0}^n (-1)^{\nu+1} \frac{d^{\nu+1}}{dx^{\nu+1}} \left( \frac{\partial f}{\partial \rho^{(\nu)}} \right)
\end{equation}
Here, we used the fact that the chemical potential $\mu$ is constant and hence its derivative is zero. By inserting Eq. \ref{R2} and Eq. \ref{R3} in Eq. \ref{DVT1D}, it becomes
\begin{equation}
	\sum_{\nu=0}^n \left[ (-1)^{\nu+1} \rho \frac{d^{\nu+1}}{dx^{\nu+1}} \left( \frac{\partial f}{\partial \rho^{(\nu)}} \right) + 2 \rho^{(\nu+1)} \frac{\partial f}{\partial \rho^{(\nu)}} \right] = \dfrac{1}{4} \rho'''
\end{equation}
Finally, the left hand side of the above equation are rearranged \cite{S03} and due to the fact that only $\rho'''$ appears on the right hand side, it is found that for any $n \geq 2$,
\begin{equation}
	\frac{\partial^2 f}{\partial \left[\rho^{(n)}\right]^2} = 0 \, .
\end{equation}
This implies that $f$ must be linear with respect of $\rho^{(n)}$. This clearly violates the non-negativity of KED and hence the dependence of $f$ on $\rho^{(n)}$ must be ruled out.}

\FH{The extension of the above proof to higher dimensions is conceptually straightforward; but more tedious mathematically. First, the differential virial theorem is extended to $D$-dimension \cite{H03}, and it becomes:
\begin{equation}
\label{DVTDD}
	\rho(x) \frac{dv}{dx_\alpha} = \dfrac{1}{4} \frac{d}{dx_\alpha} \left( \nabla^2 \rho \right) - z_\alpha
\end{equation}
where
\begin{equation}
	z_\alpha = \sum_{\beta=1}^D \left( \frac{\partial}{\partial x_\beta'} + \frac{\partial}{\partial x_\beta''} \right) \frac{\partial^2 \gamma\left( \textbf{x} + \textbf{x}' ; \textbf{x} + \textbf{x}'' \right)}{\partial x_\alpha' \partial x_\beta''} \vert_{\textbf{x}' = \textbf{x}'' = \textbf{0}}
\end{equation}
and $\gamma\left( \textbf{x} + \textbf{x}' ; \textbf{x} + \textbf{x}'' \right)$ is the one-particle density matrix. $z_\alpha$ can be decomposed into two parts as follow:
\begin{equation}
	z_\alpha = 2 \frac{dt}{dx_\alpha} + \tilde{z}_\alpha
\end{equation}
Then, we take the derivative of Euler equation in $D$-dimesion. The resulted equation is
\begin{equation}
	\frac{dv}{dx_\alpha} = \sum_{\nu=0}^n (-1)^{\nu+1} \frac{d}{dx_\alpha} \left[ \nabla^\nu \left( \frac{\partial f}{\partial \left[ \nabla^\nu \rho \right]} \right)  \right]
\end{equation}
Finally, by inserting the above equations in the differential virial equation (Eq. \ref{DVTDD}). It can be rewritten as
\begin{widetext}
\begin{equation}
	\sum_{\nu=0}^n \left[ (-1)^{\nu+1} \rho \frac{d}{dx_\alpha} \left[ \nabla^\nu \left( \frac{\partial f}{\partial \left[ \nabla^\nu \rho \right]} \right)  \right] + 2 \frac{d \left[ \nabla^\nu \rho \right]}{dx_\alpha} \frac{\partial f}{\partial \left[ \nabla^\nu \rho \right]} \right] + \tilde{z}_\alpha = \dfrac{1}{4} \frac{d}{dx_\alpha} \left( \nabla^2 \rho \right)
\end{equation}
\end{widetext}
With further arrangement and because only $\frac{d}{dx_\alpha} \left( \nabla^2 \rho \right)$ term appears on the right hand side, it can be shown that 
\begin{equation}
	\frac{\partial^2 f}{\partial \left[\nabla^n \rho\right]^2} = 0 \, .
\end{equation}
for $n \geq 2$. Again, this implies that $f$ must be linear with respect of $\nabla^n \rho$ and this clearly violates the non-negativity of KED. Thus, it must be ruled out.
}

On the other hand, the Laplacian (beside the gradient) arises naturally from the definition of kinetic energy operator \cite{A01}. However, there exists some ambiguity concerning it \cite{A01,S04,E01}. As for $t \left( \textbf{x} \right)$, unphysical pathological negativity arises if the dependence on $\nabla^2 \rho$ is of odd-power and $\nabla^2 \rho$ is negative itself. This agrees with the conclusion of Shao and Baltin eliminating dependence on $\nabla^2 \rho$. Thus, the scalar function \FH{compatible with the differential virial theorem} becomes
\begin{equation}
\label{GEFr}
	t\left( \textbf{x} \right) \equiv f \left( \rho , \nabla \rho \right) \, .
\end{equation}

In addition, it is known that the spatial extension of the density plays an important role in $t \left( \textbf{x} \right)$. Previously, this was accounted for by using either the total number of electrons $N$ or the physical coordinates, $\textbf{x}$ (please see \cite{T01} and the references within). Using $N$ --as a number-- ignores details associated with $\rho \left( \textbf{x} \right)$ while the explicit inclusion of $\textbf{x}$ violates the invariance under coordinate transformations. In this work, we incorporate the effect of the spatial extension of the density by a single measure, $r_d$, which is related to the trace of the covariance matrix $\boldsymbol{\sigma}=[\sigma_{ij}]$. It is defined conventionally as:
\begin{equation}
\label{rd1}
	r_d^2 = \dfrac{1}{D} \sum_i^D \sigma_{ii}^2 \quad,
\end{equation}
where:
\begin{equation}
\label{cmg}
	\sigma_{ij}^2 = \dfrac{\int \rho(\textbf{x})\left( x_i - \mu_i \right)\left( x_j - \mu_j \right) \textbf{dx}}{\int \rho(\textbf{x}) \, \textbf{dx}}
\end{equation}
and
\begin{equation}
\label{sav}
	\mu_i = \dfrac{\int \rho(\textbf{x}) \,  x_i \, \textbf{dx}}{\int \rho(\textbf{x}) \textbf{dx}}.
\end{equation}
By inserting Eqs. \ref{cmg}-\ref{sav} in Eq. \ref{rd1}, it becomes
\begin{equation}
\label{rd2}
	r_d^2 = \dfrac{1}{D} \dfrac{\iint \rho(\textbf{x}) \rho(\textbf{x}') \, \textbf{x} \cdot \left( \textbf{x} -\textbf{x}' \right) \, \textbf{dx} \, \textbf{dx}'}{\left[ \int \rho(\textbf{x}) \textbf{dx} \right]^2} \,.
\end{equation}
Eq. \ref{rd2} may simply be reduced to the normalized standard deviation in the 1-dimensional case. Thus, the spatial extension is derived solely from $\rho(\textbf{x})$. More interestingly, the consideration of the spatial extension leads to an additional origin of non-locality in KEDF as anticipated by many \cite{H04,K01,C03,W01,W02,P02,P03}. This shall be investigated further in a future work.

However, it is important to highlight that $r_d$ could result in some inconsistencies. The covariance matrix, $\boldsymbol{\sigma}$, (which is used to estimate $r_d$) accounts for many spatial aspects concurrently. As an example, it accounts for the spatial dispersion of a single clustered density. On the contrary, in the case of a pair of ``non-overlapping densities'', $\boldsymbol{\sigma}$ is dominated by the distance between the centers of the two densities rather than their individual spatial extensions. Thus, there is still a need to have a more consistent ``measure'' for the spatial extension of the density. This shall not alter the form of KEDF; however, $\dfrac{\delta T}{\delta \rho}$ must be modified according to the new conventionally defined ``measure''.

With all these considerations plus the scalar nature of $t \left( \textbf{x} \right)$, a possible general expansion of KEDF is given by:
\begin{equation}
\label{DExp01}
	t\left( \textbf{x} \right) = \sum_{sln} a_{sln} \frac{1}{r_d^s} \rho^{\frac{l}{D}} \left( \nabla \rho \cdot \nabla \rho \right)^{n/2},
\end{equation}
where $a_{sln}$ are the expansion coefficients and shall be determined through statistical training.

\subsection*{The limits of the sum}

Obviously, the general expansion (Eq. \ref{DExp01}) may have infinite terms as $s$, $l$, and $n$ can take any integer value. However, this is governed by physics and hence there are limits and interconnections between the three indices. By seeking the proper dimensionality of $t\left( \textbf{x} \right)$ ($L^{-D-2}$ in atomic unit, where $L$ is the dimension of the length), it is found that:

\begin{equation}
\label{lexp}
	l = \left( 2-s-n \right) + \left( 1-n \right)D \,,
\end{equation}
This permits a description of $l$ entirely in terms of $s$ and $n$ hence the expansion is reduced to:
\begin{equation}
\label{DExp02}
	t\left( \textbf{x} \right) = \sum_{sn} a_{sn} \frac{1}{r_d^s} \rho^{\frac{1}{D}\left[ \left( 2-s-n \right) + \left( 1-n \right)D \right]} \left( \nabla \rho \cdot \nabla \rho \right)^{n/2} \,.
\end{equation}

The limits of the sum are determined by other essential physical considerations. Fundamentally, since $\rho(\textbf{x})$ is finite, $n$ must be non-negative. Concerning $s$, for uniform density --which is a proper quantum mechanical case-- $r_d \rightarrow \infty$; thus, $s$ must be non-negative as well. The upper limits of $s$ and $n$ are determined by considering finite systems where $\rho\left( \textbf{x} \right)$ vanishes exponentially. For such systems, \FH{$\rho(\textbf{x})=f(\textbf{x}) \exp\left( - \kappa \sqrt{x^2} \right)$} where \FH{$\kappa$ is} positive and non-zero. By applying this restriction, it can be shown that for such systems, Eq. \ref{DExp01} is further reduced to:
\begin{equation}
\begin{split}
	t\left( \textbf{x} \right) = & \sum_{sn} \frac{a_{sn}}{r_d^s} \left[  f^{\frac{l}{D}} +  \left( \nabla f \cdot \nabla f + \kappa^2 f^2 \FH{- 2 \, \kappa \, f \frac{\nabla f \cdot \textbf{x}}{x^2} } \right)^{n/2} \right] \\ 
    & \times \exp \left\lbrace -\left( \frac{l}{D} + n \right) \FH{ \kappa \sqrt{x^2} } \right\rbrace.
\end{split}
\end{equation}

By forcing $t\left( \textbf{x} \right)$ to vanish at infinity,
\begin{equation}
	\frac{l}{D} + n > 0 \, ,
\end{equation}
and by using Eq. \ref{lexp}, we are lead to
\begin{equation}
	s+n<D+2 \,,
\end{equation}
by using Eq. \ref{lexp}. This defines the upper limits of each expansion iterator. So, the general form Eq.\ref{DExp01} is reduced to the following expansion in $D$-dimensional space:
\begin{equation}
\label{DExp}
	t\left( \textbf{x} \right) = \sum_{s=0}^{D+1} \sum_{n=0}^{D+1-s} a_{sn} \frac{1}{r_d^s} \rho^{\frac{1}{D}\left[ \left( 2-s-n \right) + \left( 1-n \right)D \right]} \left( \nabla \rho \cdot \nabla \rho \right)^{n/2} \,.
\end{equation}
This expansion captures most of the known forms of one-point KEDFs. For example, by setting, $s=n=0$, the resulted form $\rho^{(D+2)/D}$, which is simply TF KEDF in $D$-dimension. If $s=0$ and $n=2$, the resulted KEDF is vW, i.e. $\vert \nabla \rho \vert^2 / \rho$.

The axiomatic approach determines the possible KEDF forms that satisfy the essential physical requirements of dimensionality and finiteness. But, the expansion coefficients, $a_{ns}$, must be determined by other means. In this work, we will determine these $a_{ns}$ statistically in an analogous method to the machine learning approach. However, the derived functional forms are in principal universal and their coefficients must be accordingly universal. Later, we will show that this statistical approach leads to the properly known coefficients in the cases of the TF and vW limits.

\subsection*{The functional derivative}

As the application of DFT requires the minimization of the total energy as a functional of the density and subjected to some constraints, we must find the functional derivative of the kinetic energy functional:
\begin{equation}
\label{KEF}
	T[\rho\left( \textbf{x} \right)] = \int t\left( \textbf{x} \right)  \, \textbf{dx} \, .
\end{equation}
Using the general derived form of $t\left( \textbf{x} \right)$ as in Eq. \ref{DExp}, we found that:

\footnotesize
\begin{multline}
\label{DEexD}
		\dfrac{\delta T}{\delta \rho} = \sum_{s=0}^{D+1} \sum_{n=0}^{D+1-s} a_{sn} \frac{1}{r_d^s} \left\lbrace \rho^\frac{l}{D} \left( \nabla \rho \cdot \nabla \rho \right)^{n/2-2} \left[ \frac{l}{D} \left( 1 - n \right) \frac{1}{\rho} \left( \nabla \rho \cdot \nabla \rho \right)^2 - n \left( \nabla \rho \cdot \nabla \rho \right) \nabla^2 \rho - \frac{n}{2} \left( n - 2 \right)  \left[ \nabla \rho \cdot \nabla \left( \nabla \rho \cdot \nabla \rho \right) \right] \right] \right. \\
        + \left. s \int \rho(\textbf{x}')^{\frac{l}{D}} \left( \nabla \rho(\textbf{x}') \cdot \nabla \rho(\textbf{x}') \right)^{n/2} \textbf{dx}' \left[ \dfrac{1}{\int \rho(\textbf{x}') \textbf{dx}'} - \frac{1}{2} \frac{\int \rho(\textbf{x}') \left( \textbf{x}-\textbf{x}' \right) \cdot \left( \textbf{x}-\textbf{x}' \right) \textbf{dx}'}{\iint \rho(\textbf{x}') \rho(\textbf{x}'') \, \textbf{x}' \cdot \left( \textbf{x}' -\textbf{x}'' \right) \, \textbf{dx}' \, \textbf{dx}''} \right] \right\rbrace \, .
\end{multline}
\normalsize

\section*{Results and discussion for 1D cases}

In this work, we consider only non-interacting fermions in one-dimension where the occupied states are assumed to be doubly and fully occupied closed-shell. In such case, Eqs. \ref{DExp} and \ref{DEexD} are reduced to:
\begin{equation}
\label{DExp1D}
	t\left( x \right) = \sum_{s=0}^{2} \sum_{n=0}^{2-s} a_{sn} \frac{1}{r_d^s} \rho^{\left( 3-s-2n \right)} \left( \rho'^2 \right)^{n/2} \,
\end{equation}
and
\begin{multline}
\label{DEexD1D}
		\dfrac{\delta T}{\delta \rho} = \sum_{s=0}^{2} \sum_{n=0}^{2-s} a_{sn} \frac{1}{r_d^s} \left\lbrace \left(3-s-2n\right) \left( 1 - n \right) \rho^{\left(2-s-2n\right)} \left( \rho'^2 \right)^{n/2} - n \left( n - 1 \right) \, \rho^{\left( 3-s-2n \right)} \left( \rho'^2 \right)^{\left(\frac{n}{2}- 1 \right)} \rho'' \right. \\
        \left. + s \int \rho^{\left(3-s-2n\right)} \left( \rho'^2 \right)^{n/2} dx' \left[ \dfrac{1}{\int \rho(x') dx'} - \frac{1}{2} \frac{\int \rho(x') \left( x-x' \right)^2  dx'}{\iint \rho(x') \rho(x'') \, x' \left( x' -x'' \right) \, dx' \, dx''} \right] \right\rbrace
\end{multline}

\begin{figure}[ht]
\centering
\includegraphics[width=3.5in]{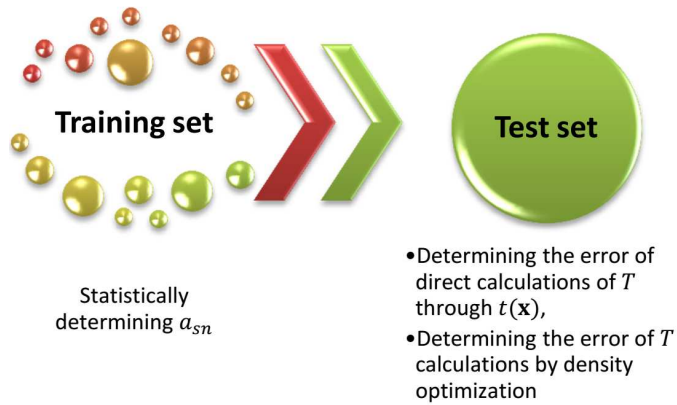}
\caption{The statistical method used to determine the expansion coefficients that are then applied in $t\left( x \right)$  to calculate the kinetic energy and to implement DFT minimization.}
\label{Prcs}
\end{figure}

The resulted expansion terms of 1D KEDF are shown in Table-1. However, we still need a mechanism to determine the expansion coefficients. In principle, these coefficients must be universal and should be determined entirely by physical consideration. For example, in the TF limit, the known coefficient for $\rho^3$ (TF model) is $a_{00}=\pi^2/24$ while $a_{02}=1/8$ is the vW coefficient. However, and aforementioned in this work, these coefficients are determined statistically by using training sets of known kinetic energies and densities for given potentials. Then, these coefficients are used to calculate the kinetic energy for a new set of "test" potentials using $t\left( x \right)$ with densities both numerically obtained by solving Schr\"odinger equation and resulting from DFT minimization. This process is represented schematically in Figure-\ref{Prcs}. It was found that the expansion coefficients converged rapidly with the size of the training set for a given number of occupied states. To guarantee consistency, training sets of 1000 potentials are used throughout this paper.

\begin{table}[H]
\centering
\begin{tabular}{cccccc}
\hline
& $s$	& $n$ &   & KEDF & \\
\hline
& 0 & 0 &   & $\rho^3$ & \\
& 0 & 1 &   & $\rho |\rho'|$ & \\
& 0 & 2 &   & $\dfrac{\left( \rho' \right)^2}{\rho}$ & \\
& 1 & 0 &   & $\dfrac{1}{r_d} \rho^2$ & \\
& 1 & 1 &   & $\dfrac{1}{r_d} |\rho'|$ & \\
& 2 & 0 &   & $\dfrac{1}{r_d^2} \rho$ & \\
\hline
\end{tabular}
\label{T1}
\caption{The expansion terms of KEDF in 1D.}
\end{table}

The considered class of potentials is the one used by Burke and coworkers which consists of three different Gaussian dips (GD) confined to a 1D box of length $L=1$ and between two infinite walls \cite{S01,L01}. This class of potentials possesses the functional form:
\begin{equation}
\label{v1}
	v(x) = - \sum_{i=1}^3 a_i \exp \left[ \dfrac{- \left( x-b_i \right)^2 }{2 c_i^2} \right] \quad,
\end{equation}
where $a_i$, $b_i$, and $c_i$ are generated randomly, and obey the following constraints: $1<a<10$, $0.4<b<0.6$, and $0.03<c<0.1$. An efficient spectral method is used to solve for the states and hence the densities with an accuracy greater than $10^{-12}$ for the exact non-interacting kinetic energy ($T_s$) \cite{FH1,FH2} .

\subsection*{$T[\rho]$ calculated using the density numerically obtained through solving the Schr\"odinger equation} 

\begin{figure}[ht]
\centering
\includegraphics[width=3.2in]{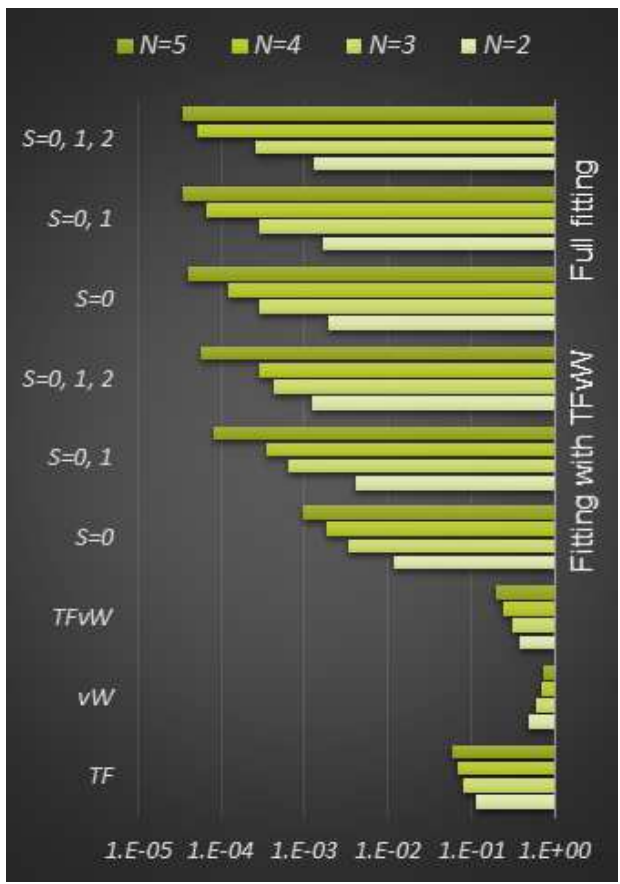}
\caption{The relative error in the non-interacting kinetic energy $T_s$. The top three piles are for full fitting where three upper limits of $s$ are assumed; namely 0, 1, and 2. The middle set of three piles are for the case where we use $a_{00}=\pi^2/24$ and $a_{02}=1/8$ to enforce the TF and vW (TFvW) limits, respectively. The last three piles are for the standard TF, vW, and TFvW models.}
\label{F02}
\end{figure}

A thousand potentials are generated and solved; the exact $T_s$ is then calculated for various numbers of occupied states ($N=2,\,3,\,4,\,5$). The Exact $T_s$ are used to find the best least square fitting for the expansion coefficients. Figure-\ref{F02} gives the performance of the expansion in terms of mean relative error ($|T_s - \int_0^L t(\rho(x)) dx|/T_s$). The top three piles are for full fitting where three upper limits of $s$ are assumed; namely 0, 1, and 2. The middle set of three piles are for the case where we use $a_{00}=\pi^2/24$ and $a_{02}=1/8$ to enforce the TF and vW (TFvW) limits, respectively. The last three piles are for the standard TF, vW, and TFvW models.

Clearly, the new expansion (Eq.\ref{DExp}) results in better performance when compared to the standard models by at least two orders of magnitude. The accuracy is improved further as the number of the occupied states increases. Also, it is clear that the consideration of the spatial extension of the density -- by using $r_d$ -- improves the estimation even further. Note that only 4 parameters (6 under the case of full fitting) are needed to very accurately estimate the non-interacting kinetic energy for 1000 potentials with four different occupied states.

\begin{figure}[ht]
\centering
\includegraphics[width=4.0in]{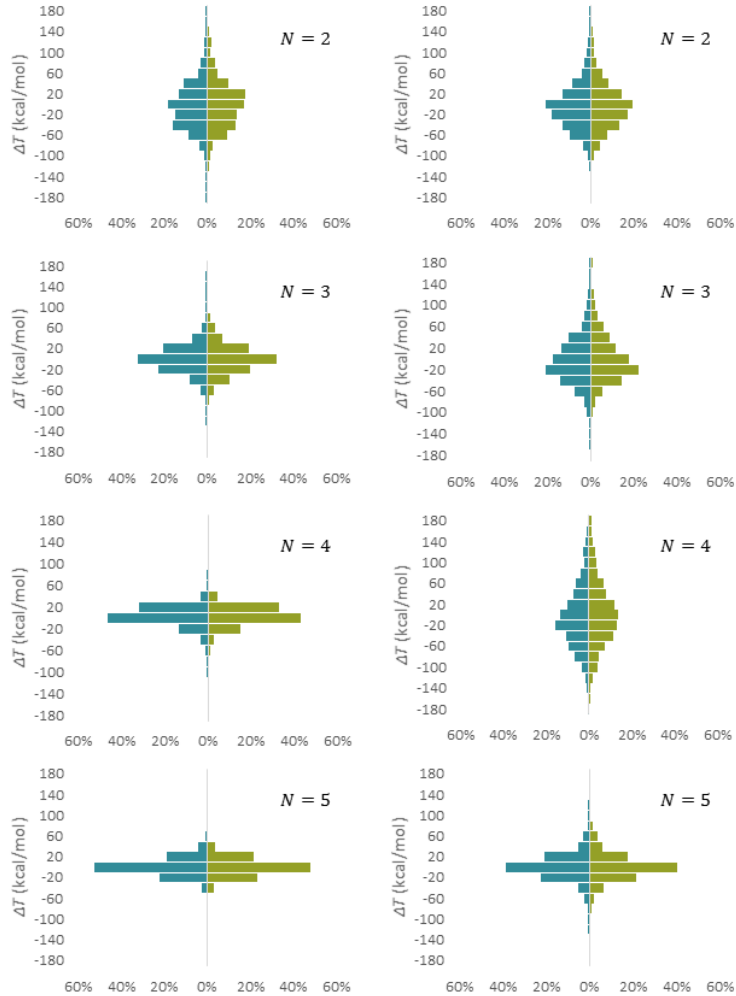}
\caption{The percentage histograms of error (in kcal/mol) of 1000 training potentials and 1000 test potentials and for different number of occupied states ($N=2,\,3,\,4,\,5$). The left column is for full fitting while the right column is for the case where $a_{00}$ and $a_{02}$ are set based on the TF and vW models. The left histograms (blue) are for the training set while the right histograms (green) are for the test set.}
\label{F03}
\end{figure}

The second analysis was designed to test the validity of the expansion. At the beginning, 1000 training potentials are used to find the expansion coefficients. These are then used to estimate $T$ directly for another 1000 test potentials for $N=2,\,3,\,4,\,5$. The results are shown in Figure-\ref{F03} where the left histograms (blue) are for the training set while the right histograms (green) are for the test set. The left column is for full fitting while the right column is for the case where $a_{00}$ and $a_{02}$ are set to $\pi^2/24$ and $1/8$ and only 4 parameters are statically determined. Here, the error is given in kcal/mol. Distinctly, the statistically trained expansion coefficients were able to estimate $T_s$ very accurately. The errors mean absolute, standard deviation, and max absolute are shown in Table-2 and are given in kcal/mol. However, they are still much beyond the chemical accuracy limit (1 kcal/mol). Furthermore and as aforementioned, the family of potentials used is the same as was used by Burke and coworkers to find KEDF with machine learning \cite{S01,L01}. In that work, they used a process containing around 100,000 empirical parameters. They achieved accuracies below 1 kcal/mol and two order of magnitudes better than what is obtained in this work. However to iterate, in this work we used only 6 parameters.

\begin{table} [t]
\centering
\begin{tabular}{ccccccc}
\hline
$N$ &  \multicolumn{3}{c}{Full Fitting} & \multicolumn{3}{c}{Fitting with TFvW}\\
 & $\overline{|\Delta T|}$ & $|\Delta T|_{std}$ & $|\Delta T|_{max}$ & $\overline{|\Delta T|}$ & $|\Delta T|_{std}$ & $|\Delta T|_{max}$ \\
\hline
2 & 40.4 & 38.0 & 283.4 & 38.9 & 37.4 & 344.3 \\
3 & 22.3 & 21.7 & 164.0 & 37.9 & 35.8 & 330.1 \\
4 & 11.9 & 10.4 & 89.0  & 53.6 & 47.2 & 268.1 \\
5 & 12.0 & 10.1 & 55.4  & 19.1 & 18.9 & 123.2 \\
\hline
\end{tabular}
\label{Table2}
\caption{The errors mean absolute ($\overline{|\Delta T|}$), standard deviation ($|\Delta T|_{std}$), and max absolute ($|\Delta T|_{max}$) in kcal/mol.}
\end{table}

\subsection*{$T[\rho]$ calculated using the density resulted from DFT minimization} 

\begin{figure*} [ht]
\centering
\includegraphics[width=6.0in]{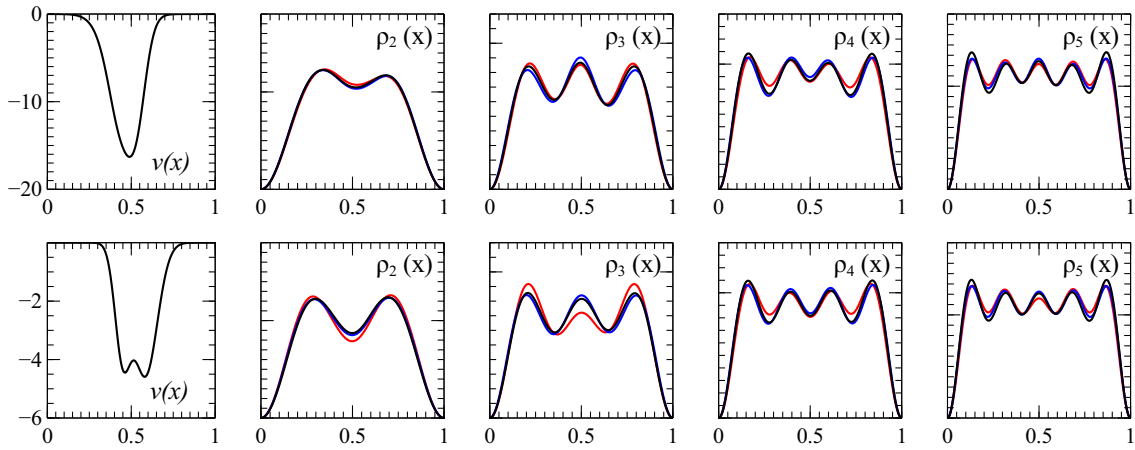}
\caption{The calculated densities from energy minimization using both full fitting (blue lines) and the fitting with forced TFvW (red lines) beside the exact density (black lines) for two randomly generated potentials and for various number of occupied states ($N=$2, 3, 4, as 5 as indicated in the top right side of each of the panels). The top set is for the first potential (shown in the top left panel) while the bottom set is for the second potential (shown in the bottom left panel).}
\label{F04}
\end{figure*}

\begin{figure}[ht]
\centering
\includegraphics[width=4.0in]{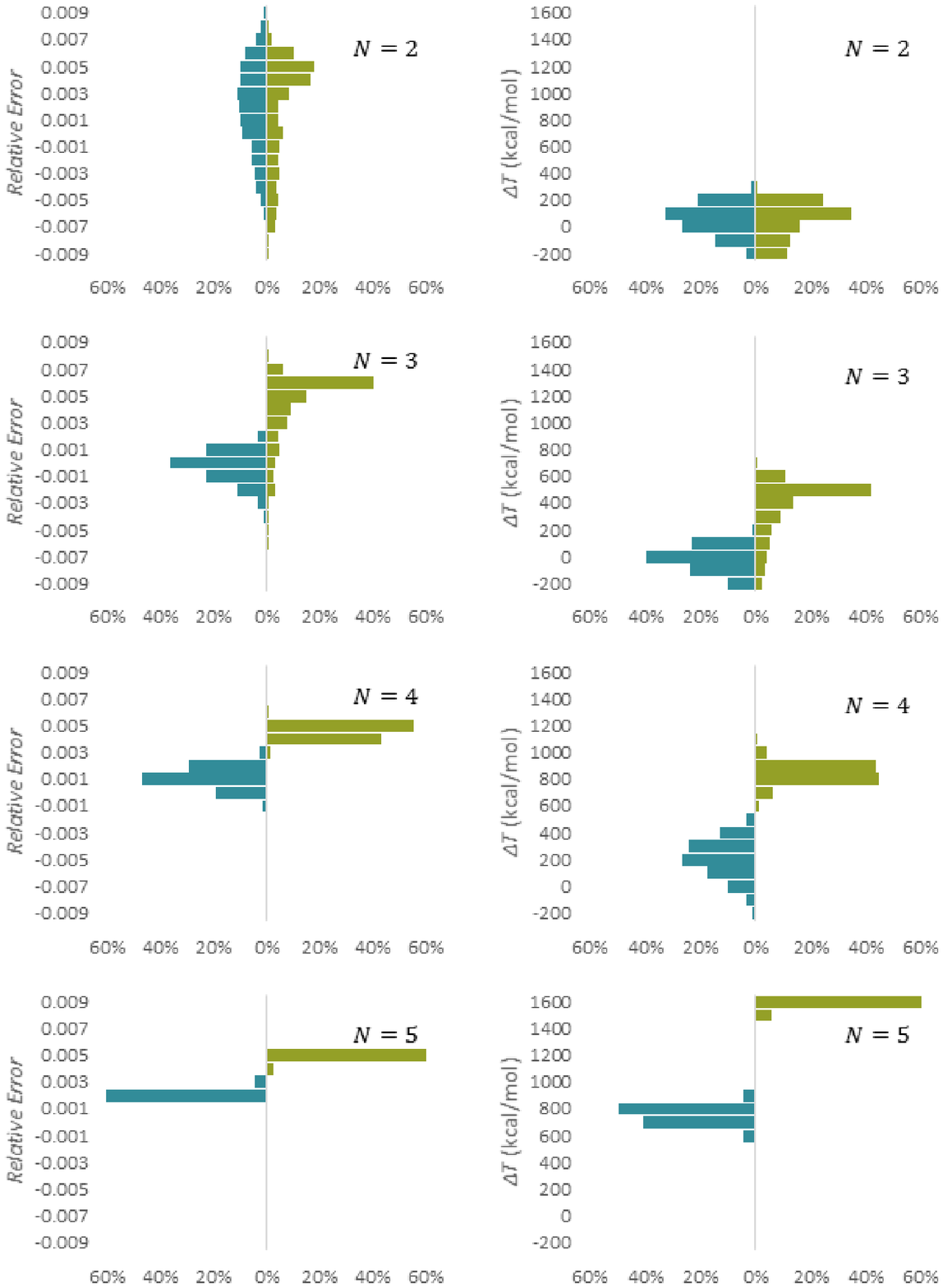}
\caption{The percentage histograms of error in relative values (left column) and in kcal/mol (right column) of 1000 potentials and for different number of occupied states ($N=2,\,3,\,4,\,5$). In each panel, the left side (blue) is for the full fitting while the right one (green) is for the fitting with TFvW imposed.}
\label{Fig5}
\end{figure}

In the previous subsection, the exact density (as calculated using high order spectral methods) are used to estimate the kinetic energy. However in DFT, it is essential to find the density that minimize the energy while maintaining the number of involved particles. There are many possible techniques \cite{S05,F02}. In this work, we use a gradient-based trust-region self-consistent method \cite{T03,F03,S06,B02}. Figure-\ref{F04} shows the calculated densities resulted from energy minimization using both full fitting and the fitting with forced TFvW beside the exact density for two randomly generated potentials and for various number of occupied states ($N=$2, 3, 4, as 5). It is clear that the suggested $t\left( x \right)$ (as tested in 1D cases) accurately produces the minimum densities. Also, the results of full fitting are better that those of the fitting with imposed TFvW parameters. However, the main outcome is its capacity to maintain the shell structure, which is challenging for the standard approximation based on TFvW models.

Energy minimization is then applied to find the densities and the kinetic energies of the 1000 systems used in the previous subsection for $N=2,\,3,\,4,\,5$. The results are shown in Figure-\ref{Fig5} where the left column shows the relative error while the right column shows the absolute error in kcal/mol. In each panel, the left side (blue) is for the full fitting while the right one (green) is for the fitting with TFvW imposed. As can be seen, the error is less than 1\% and in most cases, it is less than 0.1\%. However, this is large as absolute value. Also as observed in the previous subsection, it is clear that full fitting results in more accurate calculations as expected from the calculated densities by DFT minimization. There could be many causes for the increased accuracy beside the additional degrees of freedom allowed by full fitting. However, we believe that this could be a numerical error due to the shape optimization. This becomes more apparent for higher occupied states.

\subsection*{On the expansion coefficients}

As stated earlier, the expansion coefficients are determined statistically in this work. However, they must be universal. The only two known parameters are $a_{00}$ which equal to $\pi^2/24$ (TF KEDF) and $a_{02}$ which equal to $1/8$ (vW KEDF). vW KEDF must be equal to $T_s$ for single occupied states. Thus, a fitting for various potentials with single occupied state must get reduced to vW KEDF. This was obtained for various number of training potentials while the other expansion coefficients  in this case are fitted trivially to zero. However, they must not all vanish. Rather, the collective contributions of all the other KEDF terms -- in this limit -- must vanish.

In the other extreme, by increasing the number of the occupied states, $a_{00}$ approaches the TF limit as shown in Figure-\ref{F03}. In this analysis, the employed potentials are: a particle in a box (PiB) of a width of 1; a simple harmonic oscillator (SHO) with $\omega=1$; and three randomly generated DG potentials (Eq.\ref{v1}) but with a scaling of $L$, $b_i$, and $c_i$ by factors of 1, 2, and 3. The exactly solvable models (i.e. PiB and SHO) expeditiously reached the exact limit and the resulted $a_{00}$'s for both are hardly distinguishable.

\begin{figure}[ht]
\centering
\includegraphics[width=3.5in]{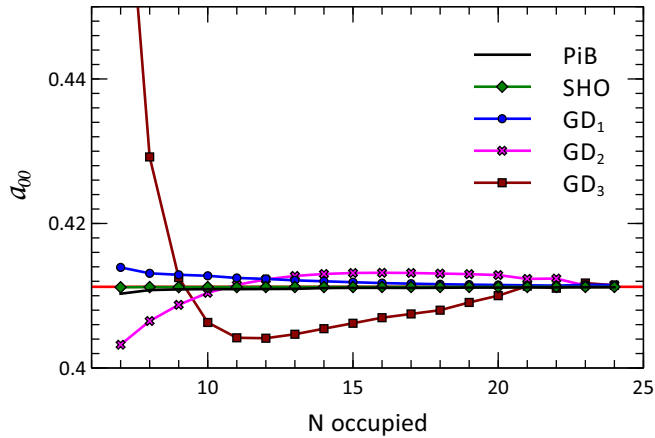}
\caption{The evolution of $a_{00}$ with the number of the occupied states for various potentials. The red line is the exact limit of $\pi^2/24$. (PiB: a particle in a box of a width of 1, SHO: a simple harmonic oscillator with, $\omega=1$, and GD$_m$: the randomly generated Gaussian dips potential (Eq.\ref{v1}) with a scaling of $L$, $b_i$, and $c_i$ by factors of 1, 2, and 3).}
\label{F05}
\end{figure}

The other expansion coefficients shall be determined by physical considerations rather than statistical fitting. This is one of many imperative goals and should permit implementing the "OF-DFT" forms to calculate the non-interacting kinetic energy in KS-DFT without solving the computationally expensive KS equations. Also, the expansion shall be implemented within a real 3D environment by application to the Coulombic potentials, where there are 15 expansion terms.

\section*{Conclusion}

The general expansion -- which was derived axiomatically through considering physical necessities -- can become a systematic approach to develop density functionals, one of the main challenges existing in the field. This shall reattain the 3D nature of DFT calculations (i.e. OF-DFT), which shall permit conduction of large-scale calculations which were inconceivable by KS-DFT. For example, Gavini and others \cite{G01} studied metallic systems with multi-million atoms by using OF-DFT. And recently, EA Carter and others analyzed a system of more than 1 million lithium atoms \cite{C04}. However, using OF-DFT with the standard KEDF for non-metallic system is inaccurate as they are known to be sufficient only for nearly-uniform density systems like metals. So, the proposed KEDF expansion should allow a systematic improvement of OF-DFT accuracy and applicability.

\nocite{*}
\bibliography{KEDF}

\end{document}